\documentclass[aps,twocolumn]{revtex4}
\begin{document}
\title{Advancing the case for  $PT$ Symmetry -- the Hamiltonian is always $PT$ Symmetric}
\author{Philip D. Mannheim}
\affiliation{Department of Physics, University of Connecticut, Storrs, CT 06269, USA.
email: philip.mannheim@uconn.edu}
\date{June 28, 2015}
\begin{abstract}
While a Hamiltonian can be both Hermitian and $PT$ symmetric, it is $PT$ symmetry that is the more general, as it can lead to real energy eigenvalues even if the Hamiltonian is not Hermitian. We discuss some specific ways in which $PT$ symmetry goes beyond Hermiticity and is more far reaching than it. We show that simply by virtue of being the generator of time translations, the Hamiltonian must always be $PT$ symmetric, regardless of whether or not it might be Hermitian. We show that the reality of the Euclidean time path integral is a necessary and sufficient condition for $PT$ symmetry of a quantum field theory, with Hermiticity only being a sufficient condition. We show that in order to construct the correct classical action needed for a path integral quantization one must impose $PT$ symmetry on each classical path, a requirement that has no counterpart in any Hermiticity condition since Hermiticity of a Hamiltonian is only definable after the quantization has been performed and the quantum Hilbert space has been constructed. With the spacetime metric being $PT$ even we show that a covariant  action must always be $PT$ symmetric. Unlike Hermiticity, $PT$ symmetry does not need to be postulated as it is derivable from Poincare invariance. Hermiticity is just a particular realization of $PT$ symmetry, one in which the eigenspectrum is real and complete.
\end{abstract}
\maketitle

\section{Introduction and Background}

Hermiticity of the Hamiltonian has been a cornerstone of quantum mechanics ever since its inception. Nonetheless, while the eigenvalues of a Hermitian Hamiltonian  are all real, Hermiticity of a Hamiltonian is only a sufficient condition for such reality. As is for instance manifested  in the matrix given in \cite{Bender2007}, viz. 
\begin{eqnarray}
M=\left(\matrix{1+i&s\cr s&1-i\cr}\right),
\label{H1}
\end{eqnarray}
we see that Hermiticity  is not a necessary condition, since even though this $M$ is not Hermitian, its eigenvalues are given by $E_{\pm}=1 \pm (s^2-1)^{1/2}$, and both of these eigenvalues are real if $s$ is real and greater than one.

A more general condition for the reality of eigenvalues has been identified by Bender and collaborators, and in a sense it is surprising since it involves an operator, time reversal $T$, that acts anti-linearly in the space of states rather than linearly, and is thus not ordinarily considered in linear algebra studies. The explicit condition that was found \cite{Bender1998,Bender1999} was that the Hamiltonian has to be $PT$ symmetric where $P$ is the parity operator, and this has engendered a large number of $PT$ studies in recent years, as described for instance in \cite{Bender2007,Special2012,Theme2013}. (In our example above, if we set $P=\sigma_1$ and $T=K$ where $K$ denotes complex conjugation we obtain $PTMT^{-1}P^{-1}=M$.) 

While $PT$ symmetry encompasses Hermiticity (Hermitian Hamiltonians can also be $PT$ symmetric), it allows for more possibilities. The matrix $M$ given in (\ref{H1}) is $PT$ symmetric for any value of  the real parameter $s$. However, if $s^2<1$ the energy eigenvalues form a complex conjugate pair. And while the energy eigenvalues would be real and degenerate at the crossover point where $s=1$, at this point the matrix becomes of non-diagonalizable Jordan-block form with $M$ only possessing one eigenvector \cite{Mannheim2013}. Neither of these possible outcomes is achievable with Hermitian Hamiltonians.

The utility in having a complex conjugate pair of energy eigenvalues is that when a state $|A\rangle$ (the state whose energy has a negative imaginary part) decays into some other state $|B\rangle$ (the one whose energy has a positive imaginary part), as the population of state $|A\rangle$ decreases that of $|B\rangle$ increases in proportion. In a $PT$-symmetric theory this interplay between the two states is found \cite{Mannheim2013} to lead to unitary time evolution. In contrast, in theories based on Hermitian Hamiltonians, to describe  a decay one essentially by hand adds a non-Hermitian term to a Hamiltonian, and again by hand chooses its sign so that only the decaying mode appears.

As regards the Jordan-Block case, we recall that in matrix theory Jordan showed that via a sequence of similarity transformations any matrix can be brought either to a diagonal form or to the Jordan canonical form in which all the eigenvalues are on the diagonal, in which the only non-zero off-diagonal elements fill one of the diagonals next to the leading diagonal, and in which all non-zero elements in the matrix are all equal to each other. To see this explicitly for our example, when $s=1$ we note that by means of  a similarity transformation we can bring $M$ to the Jordan-block  form
\begin{eqnarray}
\left(\matrix{1&0\cr i&1\cr}\right)\left(\matrix{1+i&1\cr 1&1-i\cr}\right)\left(\matrix{1&0\cr -i&1\cr}\right)=\left(\matrix{1&1\cr 0&1\cr}\right),
\label{H2}
\end{eqnarray}
with the transformed $M$ being found to only possess one eigenvector, viz.  $\widetilde{(1,0)}$, where the tilde symbol denotes transpose, even though the secular equation $|M-\lambda I|=0$ has two solutions, both with $\lambda=1$. (Since the energy eigenvalues have to share the only eigenvector available in the Jordan-block case, they must be degenerate.) Such lack of diagonalizability cannot occur for Hermitian matrices, to show that $PT$ symmetry is richer than Hermiticity. Just such lack of diagonalizability has been found to occur in fourth-order derivative theories, with the relevant Hamiltonian being shown to be $PT$ symmetric in \cite{Bender2008a} and non-diagonalizable in  \cite{Bender2008b}. Fourth-order conformal gravity theory also falls into this category \cite{Mannheim2012}, and is able to be ghost free and unitary at the quantum level because of it. 

To characterize the above set of possibilities, one should look not at the eigenvector equation $H|\psi \rangle =E|\psi \rangle$, but at the secular determinant   $f(\lambda)=|H-\lambda I|$, a determinant whose zeroes are the eigenvalues of $H$. In \cite{Bender2002} it was shown that if $H$ is $PT$ symmetric then $f(\lambda)$ is a real function of $\lambda$ (viz. in an expansion $f(\lambda)=\sum a_n\lambda^n$ all $a_n$ are real). Then in \cite{Bender2010} the converse was shown, namely if $f(\lambda)$ is a real function of $\lambda$, $H$ must have a $PT$ symmetry. If $f(\lambda)$ is a real function the eigenvalues can be real or appear in complex conjugate pairs (just as we found in our example), while if  $f(\lambda)$ is not real the condition $f(\lambda)=0$ must have a complex solution. $PT$ symmetry is thus seen to be the necessary condition for the reality of  eigenvalues, while Hermiticity is only a sufficient condition. 

In this sense $PT$ symmetry is more general than Hermiticity. However it does not represent a departure from standard quantum mechanics. Rather, it exploits a freedom that quantum mechanics has always had, a freedom that had not previously been explored. This freedom is evidenced not just in the above treatment of the eigenvalues but also in the choice of Hilbert space norm. Specifically, the eigenvector equation $i\partial_{t}|R \rangle=H|R \rangle =E|R \rangle$ only involves the kets and serves to identify right eigenvectors. Since the bra states are not specified by an equation that only involves the kets, there is some freedom in choosing them. As discussed for instance in \cite{Mannheim2013}, in general one should use not the $\langle R|R \rangle$ norm associated with the Dirac conjugate $\langle R|$ of $|R \rangle$ since $\langle R(t)|R(t) \rangle$ is not equal to $\langle R(t=0)|R(t=0) \rangle$ when the Hamiltonian is not Hermitian, with this norm then not being preserved in time. Rather, one should introduce left eigenvectors of the Hamiltonian according to $-i\partial_{t} \langle L|=\langle L|H= \langle L|E$, and use the more general norm $\langle L|R \rangle$, since for it one does have $\langle L(t)|R(t) \rangle=\langle L=0)|R(t=0) \rangle$, with this norm being preserved in time. While this norm coincides with the Dirac norm $\langle R|R \rangle$ when $H$ is Hermitian, when $H$ is not Hermitian one should use the $\langle L|R \rangle$ norm instead. As noted in \cite{Mannheim2013}, for  $PT$-symmetric Hamiltonians this norm (a norm related to the overlap of $|R\rangle$ with its $PT$ conjugate rather than with its Dirac conjugate) will be time independent and lead to unitary time evolution. Moreover, it was also shown in \cite{Mannheim2013} that $PT$ symmetry of a Hamiltonian is more than just sufficient for unitary time evolution, it is actually necessary. $PT$ symmetry of a Hamiltonian is thus the most general allowable generalization of Hermiticity that can still lead to an acceptable quantum theory. 

In \cite{Mannheim2013} a procedure was given for constructing the left eigenvectors from the right eigenvectors. Specifically, since the  energy eigenvalues of a  $PT$-symmetric Hamiltonian are either real or appear in complex conjugate pairs, it follows that in such a situation both $H$ and $H^{\dagger}$ have the same eigenspectrum. They must thus be related by a similarity transformation of the form $H^{\dagger}=VHV^{-1}$ for some in general $H$-dependent $V$, and thus obey $H^{\dagger}V=VH$. Since $\langle R|$ obeys $-i\partial_t \langle R| =\langle R| H^{\dagger}$, we find that $\langle R|V$ obeys $-i\partial_t \langle R| V=\langle R| VH$, and we can thus identify $\langle L|=\langle R|V$. Thus via the right eigenvectors and the operator $V$ one can construct the left eigenvectors. While this is not a straightforward task, as noted in \cite{Mannheim2013}, the $V$ operator must exist if the Hamiltonian is $PT$ symmetric, with a symmetry condition being  something that is much easier to check for, and thus more powerful since it guarantees that such a $V$ must exist even if one cannot explicitly construct it in closed form.

As regards the $\langle L_j(t)|R_i(t) \rangle$ norm for eigenstates $i$ and $j$ of a Hamiltonian $H$, if $|R_i(t) \rangle$ is a right eigenstate with energy eigenvalue $E_i=E_i^R+iE_i^I$, in general we can write
\begin{eqnarray}
&&\langle L_j(t)|R_i(t) \rangle=\langle R_j(t)|V|R_i(t) \rangle=
\nonumber\\
&&=\langle R_j(0)|V|R_i(0) \rangle e^{-i(E_i^R+iE_i^I)+i(E_j^R-iE_j^I)}.
\label{H3}
\end{eqnarray}
Since the norm is time independent, the only allowed non-zero norms are those that obey
\begin{eqnarray}
&&E_i^R=E_j^R,\qquad E_i^I=-E_j^I,
\label{H4}
\end{eqnarray}
i.e. precisely eigenvalues that are purely real or are in complex conjugate pairs. As noted above, in the presence of complex energy eigenvalues unitarity is maintained because the only non-zero overlap of any given right eigenvector with a complex energy eigenvalue is that with the appropriate left eigenvector with the eigenvalue needed to satisfy (\ref{H4}).

It is instructive to clarify the meaning of Hermicity. If we have a Hamiltonian $H$ with elements $H_{ij}$ in some basis, then the elements of $H^{\dagger}$ are given by $(H^{\dagger})_{ij}=H_{ji}^*$ in that same basis. Suppose that $H_{ij}=(H^{\dagger})_{ij}=H_{ji}^*$ in that basis and now apply a similarity transformation $S$ to a new basis to construct $H^{\prime}=SHS^{-1}$. In the new basis we have $[H^{\prime}]^{\dagger}=[S^{-1}]^{\dagger}H^{\dagger}S^{\dagger}=[S^{-1}]^{\dagger}HS^{\dagger}=[S^{-1}]^{\dagger}S^{-1}H^{\prime}SS^{\dagger}$. As we see,  $[H^{\prime}]^{\dagger}$ is not in general equal to $H^{\prime}$, though it would be if $S$ is unitary. The reason for this is that while a unitary transformation preserves orthogonality of the basis vectors, a general similarity transformation does not as it is a transformation to a skew basis. The statement that $(H^{\dagger})_{ij}=H_{ji}^*$ is thus basis dependent. Thus to say that a Hamiltonian is Hermitian is to say that one can find a basis in which $H_{ij}$ is equal to $H_{ji}^*$, with the basis-independent statement being that the eigenvalues of a Hermitian operator are all real and the eigenvectors are complete. (For our example in (\ref{H1}) for instance, when $s^2>1$, on setting $s=\cosh\alpha$, we can write
\begin{eqnarray}
&&(A\sigma_0+B\sigma_2)(\sigma_0+i\sigma_3+\cosh\alpha\sigma_1)(A\sigma_0-B\sigma_2)
\nonumber\\
&&=\sigma_0+\sinh\alpha\sigma_1,
\label{H5}
\end{eqnarray}
where 
\begin{eqnarray}
A=\left(\frac{\coth\alpha+1}{2}\right)^{1/2},\qquad
B=\left(\frac{\coth\alpha -1}{2}\right)^{1/2},
\label{H6}
\end{eqnarray}
to thus bring $M$ to a Hermitian form.) When $s^2>1$ the matrix $M$ of (\ref{H1}) is thus Hermitian in disguise, with the utility of $PT$ symmetry being that one can learn about properties of the eigenvalues of a matrix by testing for its $PT$ symmetry, even if one cannot actually determine those eigenvalues in closed form. Moreover, unlike the condition $H^{\dagger}=H$, the relation $[H,PT]=0$ is not basis dependent, though, as noted in \cite{Mannheim2013},  under a similarity transformation it would be the transformed $P$ and $T$ that would obey $[H^{\prime},P^{\prime}T^{\prime}]=0$ \cite{footnote1}. In addition we note that suppose we are given some general $H$ in some general basis. It may or may not be Hermitian in disguise, and to check we would either have to construct an explicit similarity transformation that brings $H$ to a  basis in which $H_{ij}=H_{ji}^*$, or solve the eigenvector equation to get all the eigenvectors and eigenvalues, show that all the energy eigenvalues are real and show that the eigenvectors are complete. Alternatively, we could check to see if the Hamiltonian has a $PT$ symmetry (possible written as the product of some generalized linear operator times some generalized anti-linear one), and if this proves not to be the case we can immediately conclude that the Hamiltonian is not Hermitian in disguise. $PT$ symmetry is thus a necessary condition for Hermiticity, one that requires no need to bring $H$ to a Hermitian form \cite{footnote2}, \cite{footnote3}.

With there being two options for a Hamiltonian ($PT$ symmetry or Hermiticity), one has to ask what determines which one is to be used. To this end we look at the implications of  the Poincare algebra, and in Sec. II we show that simply by virtue of being the generator of time translations the Hamiltonian (and equally the energy-momentum tensor from which it is built according to $H=\int d^3x T^{00}$) must be $PT$ symmetric, regardless of whether or not it is Hermitian. $PT$ symmetry thus outperforms Hermiticity in this regard, with $PT$ symmetry having a direct connection to spacetime symmetries that Hermiticity does not. Moreover, in a sense this is to be anticipated since $P$ and $T$ are themselves rooted in spacetime, with it being the $PT$ product rather than individual $P$ or $T$ transformations that acts equally on all four spacetime coordinates by effecting $x^{\mu}\rightarrow -x^{\mu}$. Thus the Hamiltonian of any covariant theory must be $PT$ symmetric. Then, since it is known that this symmetry is violated in weak interactions (weak interactions conserve $CPT$ and violate $C$ where $C$ is charge conjugation), weak interaction $PT$ symmetry must be spontaneously broken. (As noted in \cite{Mannheim2014}, giving a right- or a left-handed neutrino Majorana mass operator $\tilde{\psi}(1\pm\gamma^5)i\gamma^2\gamma^0(1\pm \gamma^5)\psi$ a vacuum expectation value would spontaneously break $PT$.) 

While the work of \cite{Bender2002} and \cite{Bender2010} showed that $PT$ symmetry is a necessary and sufficient condition for the reality of the secular equation $f(\lambda)=|H-\lambda I|$, as such, the proof only applied to finite-dimensional systems such as matrices. In Sec. III we give the extension to the infinite-dimensional case by showing that   $PT$ symmetry is a necessary and sufficient condition for the reality of the field-theoretic Euclidean time path integral,  while Hermiticity is only a sufficient condition for such reality.

In quantizing a physical system one can work directly with quantum operators acting on a Hilbert space and impose canonical commutation relations for the operators, a q-number approach, or one can quantize using Feynman path integrals, a purely c-number approach. In constructing the appropriate classical action needed for the path integral approach, one ordinarily builds the action out of real quantities, because real quantities are the eigenvalues of Hermitian  quantum operators. However, as we show in Sec. IV, this is inadequate in certain cases, and particularly so in minimally coupled electrodynamics (while $\partial_{\mu}-A_{\mu}$  is real, it is only $i\partial_{\mu}-A_{\mu}$ that can be Hermitian in the quantum case), with the correct classical action being constructed by requiring that it be $PT$ symmetric instead ($i\partial_{\mu}$ and $A_{\mu}$ are both $PT$ even in both the classical and the quantum cases). In Sec. IV we buttress this result by showing that since both $T^{\mu\nu}$ and the spacetime metric $g^{\mu\nu}$ are $PT$ even, any covariant action must be $PT$ even too.

As constructed in quantum mechanics, to show that an operator such as the momentum operator  $i\partial_x$ (or the Hamiltonian that is built out of it) acts as a Hermitian operator in the space of wave functions, one has to integrate by parts and be able to throw away spatially asymptotic surface terms. $PT$ symmetry generalizes this notion by allowing for the possibility that one may have to rotate into the complex plane in order to find so-called Stokes wedges in which one can throw surface terms away \cite{Bender2007} when it is not possible to do so on the real axis. A typical example is the divergent Gaussian $\exp(x^2)$. It is not normalizable on the real $x$-axis, but is normalizable on the imaginary $x$-axis, and would be of relevance if the momentum operator $p$ were to be anti-Hermitian rather than Hermitian, and thus represented by $\partial_{x}$, with the $[x,p]=i$ commutator being realized as $[-ix,\partial_x]=i$. In fact, until one has looked at asymptotic boundary conditions, one cannot determine whether an operator is self-adjoint or not, since such self-adjointness is determined not by the operator itself but by the space of states on which it acts. The art of $PT$-symmetric theories then is the art of determining in which domain in the relevant complex plane a theory is well-behaved asymptotically, with many examples being provided in \cite{Bender2007}. However, one has to ask what happens to $PT$ symmetry as one does continue into the complex plane. In Sec. V we show that as one makes such a continuation both the $PT$ operator and the Hamiltonian transform so that their commutation relation is preserved, just as in \cite{footnote1}.

\section{The Hamiltonian is Always $PT$ Symmetric}

Consider a flat Minkowski spacetime  with metric $\eta_{\mu\nu}$ and action $I_{\rm M}=\int d^4x{{\cal L}}_{\rm M}$, where ${{\cal L}}_{\rm M}$ is a function of the various quantum matter fields of the theory. On generalizing this action to curved spacetime, one obtains a metric $g_{\mu\nu}$ and action $I_{\rm M}=\int d^4x(-g)^{1/2} {{\cal L}}_{\rm M}$, where in ${{\cal L}}_{\rm M}$ all derivatives are now  covariant ones. Functional variation of this action with respect to $g_{\mu\nu}$ yields a matter field energy-momentum tensor $T^{\mu\nu}=2(-g)^{-1/2}\delta I_{\rm M}/\delta g_{\mu\nu}$. As constructed, $T^{\mu\nu}$ is symmetric in its indices, and in solutions to the field equations associated with the variation of the action $I_{\rm M}$ with respect to the fields in ${{\cal L}}_{\rm M}$,  $T^{\mu\nu}$ is covariantly conserved according to  $\nabla_{\mu}T^{\mu\nu}=0$. On having identified an appropriate $T^{\mu\nu}$, on then returning back to the flat Minkowski case where $\partial_{\mu}T^{\mu\nu}=0$, one introduces the Lorentz generator densities
\begin{eqnarray}
M^{\sigma \mu\nu}= T^{\sigma\mu}x^{\nu}-T^{\sigma\nu}x^{\mu},
\label{H7}
\end{eqnarray}
and finds that the Lorentz generator densities obey $\partial_{\sigma}M^{\sigma \mu\nu}=0$. Given this conservation condition we can set
\begin{eqnarray}
\frac{d}{dt}\int d^3x M^{0 \mu\nu}= -\int d^3x \frac{\partial}{\partial x_i}M^{i\mu\nu}={\rm surface~term},
\label{H8}
\end{eqnarray}
where $i=1,2,3$. Consequently, if the matter fields  that obey the matter field equations of motion are sufficiently damped at spatial infinity so as to cause the surface term to vanish, the six Lorentz generators $M^{\mu\nu}=\int d^3x M^{0 \mu\nu}$ will then be  time-independent constants of the motion. And as such, since they are Lorentz generators they obey the Lorentz algebra:
\begin{eqnarray}
[M_{\mu\nu},M_{\rho\sigma}]&=&i(-\eta_{\mu\rho}M_{\nu\sigma}+\eta_{\nu\rho}M_{\mu\sigma}
\nonumber\\
&-&\eta_{\mu\sigma}M_{\rho\nu}+\eta_{\nu\sigma}M_{\rho\mu}).
\label{H9}
\end{eqnarray}

Now while we could apply separate $P$ or $T$ transformations to the Lorentz generators, we note that classical $PT$ transformations have a direct connection to the Lorentz group since all four components of $x^{\mu}$ are treated equally, with the $PT$ transformation that takes $x_{\mu}$ to $-x_{\mu}$ corresponding to a sequence of three complex Lorentz transformations $x^{\prime}=x\cosh\xi +t\sinh \xi $, $y^{\prime}=y\cosh \xi +t\sinh \xi $, $z^{\prime}=z\cosh \xi +t\sinh \xi $, each with a complex boost angle $\xi=i\pi$. A classical $PT$ transformation is thus equivalent to a complex Lorentz transformation.

In the quantum case, if we apply a $PT$ transformation to the quantum fields, $PT$ does not act on the coordinates themselves but transforms the fields at $x^{\mu}$ into fields at $-x^{\mu}$ with appropriate phases, while the anti-linear nature of time reversal complex conjugates factors of $i$. On applying $PT$ to the Lorentz generator densities we obtain $PTM^{\sigma \mu\nu}(x^{\lambda})T^{-1}P^{-1}=\theta M^{\sigma \mu\nu}(-x^{\lambda})$ where $\theta$ is a real phase. On applying this same transformation to the Lorentz algebra given in (\ref{H9}), because the factor $i$ in it is conjugated, we find that the phase $\theta$ is given by $\theta=-1$.

If we apply a $PT$ transformation  to $T^{\mu\nu}$ we obtain $PT^{\mu\nu}(x^{\lambda})T^{-1}P^{-1}=\phi T^{\mu\nu}(-x^{\lambda})$, where $\phi$ is a real phase. Given this transformation, on applying a $PT$ transformation to (\ref{H7}) we obtain 
\begin{eqnarray}
&&\theta M^{\sigma \mu\nu}(-x^{\lambda})=\phi [T^{\sigma\mu}(-x^{\lambda})x^{\nu}-T^{\sigma\nu}(-x^{\lambda})x^{\mu}]
\nonumber\\
&&=-\phi [T^{\sigma\mu}(-x^{\lambda}))(-x^{\nu})-T^{\sigma\nu}(-x^{\lambda})(-x^{\mu})]
\nonumber\\
&&=-\phi M^{\sigma \mu\nu}(-x^{\lambda}).
\label{H10}
\end{eqnarray}
We thus infer that $\phi=-\theta$, and thus that $\phi=1$. Consequently the matter field  $T^{\mu\nu}$ is $PT$ even. Thus from the structure of the Lorentz algebra alone we infer that the matter field  $T^{\mu\nu}$ is $PT$ even.

To define momentum generators we set $P^{\mu}=\int d^3x T^{0\mu}$, and from the conservation of $T^{\mu\nu}$ and the vanishing of the fields at spatial infinity, we find that the Hamiltonian $H=\int d^3x T^{00}$ is a time-independent constant of the motion, and thus equal to its value at $t=0$.  Since a $PT$ transformation on the fields does not affect $\int d^3x$, we obtain
\begin{eqnarray}
PTHT^{-1}P^{-1}&=&\int d^3x T^{00}(t=0,-x^i)
\nonumber\\
&=&\int d^3x T^{00}(t=0,x^i)=H.
\label{H11}
\end{eqnarray}
We thus conclude that the Hamiltonian is necessarily $PT$ symmetric.  Thus simply by virtue of being the generator of time translations the Hamiltonian is automatically $PT$ symmetric. And it  is $PT$ symmetric regardless of whether or not it might be Hermitian, and regardless of whether or not it might even be diagonalizable. Moreover, since the only allowable non-relativistic theories in physics are those that  are the non-relativistic limits of covariant ones (even if a system is non-relativistic the observer is allowed to be relativistic),  the only allowable non-relativistic Hamiltonians must be $PT$ symmetric too.

In choosing an appropriate energy momentum tensor we recall that there is a better choice than the canonically defined one used above. While there is no need to make any adjustment for fermions or gauge bosons \cite{Callan1970}, in the scalar field case rather than use the $T^{\mu\nu}=\partial_{\mu}\phi\partial_{\nu}\phi-g_{\mu\nu}{{\cal L}}$ associated with the variation of the action $-\int d^4x {{\cal L}}$ with ${{\cal L}}=(1/2)\partial_{\mu}\phi\partial^{\mu}\phi-V(\phi)$, Callan, Coleman, and Jackiw \cite{Callan1970} noted that one should augment this $T^{\mu\nu}$ with an extra term $\Delta T^{\mu\nu}=-(1/3)\partial_{\mu}\phi\partial_{\nu}\phi -(1/3)\phi\partial_{\mu}\partial_{\nu}\phi+(1/3)\eta_{\mu\nu}\partial_{\alpha}\phi\partial^{\alpha}\phi +(1/3)\eta_{\mu\nu}\phi \partial_{\alpha}\partial^{\alpha}\phi$, to give a total  $\theta^{\mu\nu}=T^{\mu\nu}+\Delta T^{\mu\nu}$. This $\theta^{\mu\nu}$ has two interesting properties. First, its matrix elements are much better behaved in the ultraviolet than those of $T^{\mu\nu}$ itself. And second, with this $\theta^{\mu\nu}$ one can construct the four-vector  $x_{\nu}\theta^{\mu\nu}$ and the rank two tensor $(\eta^{\mu\lambda}x_{\alpha}x^{\alpha}-2x^{\mu}x^{\lambda})\theta^{\nu}_{\phantom{\nu}\lambda}$, with these two quantities serving as the densities for scale and conformal transformations. Since both $T^{\mu\nu}$ and $\Delta T^{\mu\nu}$ are $PT$ even, $\theta^{\mu\nu}$ is $PT$ even too. Consequently, just as before, the generalized $H=\int d^3x \theta^{00}$ is equally $PT$ symmetric.

The $\Delta T^{\mu\nu}$ term can be generated by adding on to ${{\cal L}}$ the term $-(1/12)\phi^2R^{\alpha}_{\phantom{\alpha}\alpha}$, and then switching off the curvature after  functional variation with respect to the metric. However this addition is associated with a conformally coupled scalar field, with the quantity $(1/2)\partial_{\mu}\phi\partial^{\mu}\phi-(1/12)\phi^2R^{\alpha}_{\phantom{\alpha}\alpha}$ being invariant under a local conformal transformation of the form $\phi(x)\rightarrow e^{-\alpha(x)}\phi(x)$, $g_{\mu\nu}(x)\rightarrow e^{2\alpha(x)}g_{\mu\nu}(x)$. Now as had been pointed out in \cite{Mannheim2014}, a $PT$ transformation corresponds to a specific transformation of the conformal group $SO(4,2)$, the full symmetry of the light cone [cf. $PT\psi(t,\mathbf{x})T^{-1}P^{-1}=-\gamma^2\gamma^5\psi(-t,-\mathbf{x})$ for fermions where the axial currents are generators of $SU(2,2)$, the covering group of $SO(4,2)$]. Thus in the use of $H=\int d^3x \theta^{00}$, we again find a connection between $PT$ symmetry and conformal symmetry.

\section{$PT$ Symmetry and Euclidean Time Path Integrals}

If we have right and left eigenstates $|\psi_1 \rangle$ and $\langle \psi_2 |$ of a Hamiltonian with eigenvalues $E_1$ and $E_2$ according to $H|\psi_1\rangle=E_1|\psi_1 \rangle$, $\langle \psi_2|H=\langle \psi_2|E_2$, then on applying the $PT$ operator we obtain
\begin{eqnarray}
[H,PT]|\psi_1 \rangle&=&(H-E_1^*)PT|\psi_1 \rangle,
\nonumber\\
\langle \psi_2 |[H,PT]&=&\langle \psi_2 | PT(E_2^*-H).
\label{H12}
\end{eqnarray}
From (\ref{H12}) we see that if $[H,PT]=0$, then for every eigenstate $|\psi_1 \rangle$ with eigenvalue $E_1$ there exists an eigenstate   $PT|\psi_1 \rangle$ with eigenvalue $E_1^*$ (any one of which could be real of course), and likewise for $\langle \psi_2 |$. Thus, as noted above, when the Hamiltonian is $PT$ symmetric the eigenvalues can be real or appear in complex conjugate pairs. To establish the converse, we note that (\ref{H12}) has to hold not just for one of its eigenstates but for all of them. Then if some of the energy eigenvalues of $H$ are real and the rest appear in complex conjugate pairs, and if the set  of all eigenstates is complete (viz. the non-Jordan-block case),  we conclude that we can set $[H,PT]=0$ as an operator identity. We thus conclude that if some or all of the energy eigenvalues are real and any others appear in complex conjugate pairs, the Hamiltonian must be $PT$ symmetric. The utility of this analysis is that unlike the discussion of $f(\lambda)=|H-\lambda I|$ given earlier, the implications of (\ref{H12}) hold for infinite-dimensional operators and not just for finite-dimensional ones. 

In the Jordan-block case the completeness discussion has to include non-stationary solutions to the Schr\"{o}dinger equation as well as the stationary ones \cite{Bender2008b}.  To be specific, consider two typical eigenstates of an eigenvector problem of the form $\exp(-i(\omega +\epsilon )t)$ and $\exp(-i(\omega -\epsilon )t)$. On letting $\epsilon$ go to zero the two eigenstates collapse on to just the one $[\exp(-i(\omega +\epsilon )t)+\exp(-i(\omega -\epsilon )t)]/2\rightarrow \exp(-i\omega t)$,  and an eigenstate is lost. However two solutions cannot turn into one, and there thus has to be a second solution. This second solution is found by combining the two eigenstate solutions with a singular weight $1/\epsilon$ to yield $[\exp(-i(\omega +\epsilon )t)-\exp(-i(\omega -\epsilon )t)]/2\epsilon \rightarrow -it\exp(-i\omega t)$. Because of its power behavior in $t$ this second solution is not an eigenstate of $i\partial_t$. Since both of the two eigenstate solutions were independent solutions to the time-dependent Schr\"{o}dinger equation before we took the limit (since they were solutions to $i\partial_t\psi(t)=H\psi(t)=E\psi(t)$), the two linear combinations remain solutions to the time-dependent Schr\"{o}dinger equation after the limit is taken, with there thus being no loss of completeness. In analog to (\ref{H12}), starting from $i\partial_t\psi(t)=H\psi(t)$ with time-independent $H$ we can write
\begin{eqnarray}
[H,PT]\psi(t)&=&HA\psi^*(-t)+i\frac{\partial}{\partial t}A\psi^*(-t)
\nonumber\\
&=&HA\psi^*(-t)-i\frac{\partial}{\partial (-t)}A\psi^*(-t),
\label{H13}
\end{eqnarray}
where we have set $PT=AK$ where $A$ is a time-independent linear operator and $K$ is complex conjugation. From (\ref{H13}) we see that if $[H,PT]=0$, then for every $\psi(t)$ that satisfies the time-dependent Schr\"{o}dinger equation $A\psi^*(t)$ will do so too. Similarly, if for every $\psi(t)$ that satisfies the time-dependent Schr\"{o}dinger equation there is an $A\psi^*(t)$ that does so too, then we can identify $[H,PT]=0$ as an operator identity. 

We can also constrain the structure of solutions to the theory in the complex conjugate pair case. Even though all solutions are energy eigenstates in this case, nonetheless  such stationary solutions to the time-independent Schr\"{o}dinger equation $H\psi(t)=E\psi(t)$ also obey the time-dependent Schr\"{o}dinger equation $i\partial_t\psi(t)=H\psi(t)$ as well, and thus obey (\ref{H13}). On applying (\ref{H13}), we see that any complex conjugate pair of energy eigenvectors of a $PT$ symmetric Hamiltonian have, up to an $A$-dependent transformation,  complex conjugate wave functions.  

To apply the above analysis to path integrals, consider the generic path integral  $\int {{\cal D}}[\phi]\exp(iS)$ with classical action $S=\int d^4x {{\cal L}}$, as integrated over the paths of some generic field $\phi(\vec{x},t)$ between end points $\phi(\vec{x}=0,t=0)$ and $\phi(\vec{x}=0,t)$. In theories in which the Hamiltonian is Hermitian, this path integral represents the $t>0$ two-point function
\begin{eqnarray}
\langle \Omega|\phi(0,t)\phi(0,0)|\Omega\rangle e^{-iE_0t}
=\int_{\phi(0,0)}^{\phi(0,t)} {{\cal D}}[\phi]\exp(iS),~~
\label{H14}
\end{eqnarray}
where $E_0$ is the energy of the ground state $|\Omega\rangle$.
On introducing the time evolution operator, using the completeness relation  $H=\sum_n|n\rangle E_{n}\langle n|$,  and taking $\phi(\vec{x},t)$ to be Hermitian, evaluation of the two-point function yields 
\begin{eqnarray}
&&\langle \Omega|\phi(0,t)\phi(0,0)|\Omega\rangle e^{-iE_0t}
\nonumber\\
&&=\langle \Omega|e^{iHt}\phi(0,0)e^{-iHt}\phi(0,0)|\Omega\rangle e^{-iE_0t}
\nonumber\\
&&=\sum_n\langle \Omega |\phi(0,0)|n\rangle \langle n|\phi(0,0) |\Omega\rangle e^{-iE_nt}
\nonumber\\
&&=\sum_n|\langle \Omega |\phi(0,0)|n\rangle |^2 e^{-iE_nt}.
\label{H15}
\end{eqnarray}
In arriving at this result we have identified $\langle n|\phi(0,0) |\Omega\rangle$ as the complex conjugate of $\langle \Omega |\phi(0,0)|n\rangle$. Such an identification can immediately be made if the states $|n\rangle$ are also eigenstates of a Hermitian $\phi(0,0)$, except for the fact that they actually cannot be since $[\phi,H]=i\partial_t\phi$ is not equal to zero. Nonetheless, in its own eigenbasis we can set  $\phi=\sum_{\alpha}|\alpha\rangle \phi_{\alpha}\langle\alpha|$, where the $\phi_{\alpha}$ are real. Consequently, we can set
\begin{eqnarray}
&&\langle \Omega|\phi(0,0)|n\rangle =\sum_{\alpha}\langle \Omega |\alpha\rangle \phi_{\alpha}\langle\alpha |n\rangle,
\nonumber\\
&&\langle n|\phi(0,0)|\Omega\rangle =\sum_{\alpha}\langle n |\alpha\rangle \phi_{\alpha}\langle\alpha |\Omega\rangle
\nonumber\\
&&=\sum_{\alpha}\langle\alpha |n\rangle^*\phi_{\alpha} \langle \Omega |\alpha\rangle^* =\langle \Omega|\phi(0,0)|n\rangle^*,
\label{H16}
\end{eqnarray}
from which the last equality in (\ref{H15}) then follows after all.

If we now substitute the Euclidean time $\tau=it$ in  (\ref{H15}) we obtain 
\begin{eqnarray}
&&\langle \Omega|\phi(0,t)\phi(0,0)|\Omega\rangle e^{-iE_0t}
\nonumber\\
&&=\sum_n|\langle \Omega |\phi(0,0)|n\rangle |^2 e^{-E_n\tau}.
\label{H17}
\end{eqnarray}
In Euclidean time this expression is real since all the eigenvalues of a Hermitian Hamiltonian are real, and is convergent at large $\tau$ if all the $E_n$ are greater or equal to zero. Also, its expansion at large $\tau$ is dominated by $E_0$, with the next to leading term being given by next lowest energy $E_1$ and so on. Finally, in order for (\ref{H17})  to be describable by a Euclidean time path integral with convergent exponentials, as per \cite{footnote4} we would need $iS=i\int dt d^3x {{\cal L}}=\int d\tau d^3x {{\cal L}}$ to be real and negative definite on every path. 

We can obtain an analogous outcome when the Hamiltonian is not Hermitian, and as we now show, it will precisely be $PT$ symmetry that will achieve it for us. As  described earlier, in general we must distinguish between left and right eigenvectors, and so in general  the path integral will represent $\langle \Omega_{\rm L}|\phi(0,t)\phi(0,0)|\Omega_{\rm R}\rangle e^{-iE_0t}$. Now in the event that the left eigenvectors are not the Dirac conjugates of the right eigenvectors of $H$, the general completeness and orthogonality relations (in the non-Jordan-block case) are given by \cite{Mannheim2013} $\sum_n|R_{n}\rangle\langle L_{n}|=\sum_n|L_{n}\rangle\langle R_{n}|=I$, $\langle L_{n}|R_{m}\rangle=\langle R_{m}|L_{n}\rangle=\delta (n,m)$, while the spectral decomposition of the Hamiltonian is given by $H=\sum_n|R_{n}\rangle E_{n}\langle L_{n}|$. Consequently, we can set
\begin{eqnarray}
&&\langle \Omega_{L}|\phi(0,t)\phi(0,0)|\Omega_{R}\rangle e^{-iE_0t}
\nonumber\\
&&=\sum_n\langle \Omega_{L}|\phi(0,0)|R_{n}\rangle e^{-iE_{n}t}\langle L_{n}|\phi(0,0)|\Omega_{R}\rangle.
\label{H18}
\end{eqnarray}

To analyze this expression we will need to determine the matrix elements of $\phi(0,0)$. To use Hermiticity for $\phi(0,0)$ is complicated and potentially not fruitful. Specifically, if we insert  $\phi=\sum_{\alpha}|\alpha\rangle \phi_{\alpha}\langle\alpha|$ in the various matrix elements  of interest, on recalling that  $\langle L|=\langle R|V$, we obtain 
\begin{eqnarray}
&&\langle \Omega_{L}|\phi(0,0)|R_{n}\rangle =\sum_{\alpha}
\langle \Omega_R |V|\alpha\rangle \phi_{\alpha}\langle\alpha |R_n\rangle,
\nonumber\\
&&\langle L_{n}|\phi(0,0)|\Omega_{R}\rangle =\sum_{\alpha}\langle R_n |V|\alpha\rangle \phi_{\alpha}\langle\alpha |\Omega_R\rangle
\nonumber\\
&&=\sum_{\alpha}\langle\alpha |V^{\dagger}|R_n\rangle^*\phi_{\alpha} \langle \Omega_R |\alpha\rangle^*. 
\label{H19}
\end{eqnarray}
This last expression is not only not necessarily equal to $\langle \Omega_L|\phi(0,0)|R_n\rangle^*$, it does not even appear to be related to it.

To be able to obtain a quantity that does involve the needed complex conjugate, we note that  as well as being Hermitian, as a neutral scalar field, $\phi(0,0)$ is $PT$ even. Its $PT $ transformation properties are straightforward since we can write everything in the left/right energy eigenvector basis (as noted in \cite{footnote1} the relation $[PT,\phi]=0$ and thus $PT\phi T^{-1}P^{-1}=\phi$ are basis independent). On applying a $PT$ transformation and recalling that $P^2=1$, $T^2=1$ as per \cite{footnote1}, we obtain  
\begin{eqnarray}
\phi&=&\sum_{i,j}|R_i \rangle \phi_{ij}\langle L_j|=PT\phi T^{-1}P^{-1}
\nonumber\\
&=&PT\phi TP=\sum_{i,j}PT|R_i \rangle \phi^*_{ij}\langle L_j|TP.
\label{H20}
\end{eqnarray}
For energy eigenvalues that are real we have $PT|R_i \rangle=|R_i \rangle$, $\langle L_j|TP=\langle L_j|$, with $PT\phi TP=\phi$  thus yielding
\begin{eqnarray}
\phi_{ij}=\phi^*_{ij},\qquad \langle L_i|\phi |R_j \rangle =\phi_{ij}.
\label{H21}
\end{eqnarray}
Thus we can set
\begin{eqnarray}
\langle \Omega_{\rm L}|\phi(0,t)\phi(0,0)|\Omega_{\rm R}\rangle e^{-iE_0t}=\sum_n \phi_{0n}\phi_{n0} e^{-iE_{n}t}.
\label{H22}
\end{eqnarray}
With $\phi_{0n}$ and $\phi_{n0}$ both being real, this expression is completely real when the time is Euclidean. Thus in the real eigenvalue sector of a  $PT$-symmetric theory, the Euclidean time two-point function and the Euclidean time path integral are completely real. Since they both are completely real, we confirm that the form $\langle \Omega_{\rm L}|\phi(0,t)\phi(0,0)|\Omega_{\rm R}\rangle$ is indeed the correct $PT$-symmetry generalization of the Hermitian theory form $\langle \Omega|\phi(0,t)\phi(0,0)|\Omega\rangle$ used in (\ref{H14}) above.

In the event that energy eigenvalues appear in complex conjugate pairs, we have two cases to consider, namely cases in which there are also real eigenvalues, and cases in which all eigenvalues are in complex conjugate pairs. In both the cases we shall sequence the energy eigenvalues in order of increasing real parts of the energy eigenvalues. Moreover, in cases where there are both real and complex energy eigenvalues we shall take the lowest one to have a purely real energy. 

For energy eigenvalues that are in complex conjugate pairs according to $E_{\pm}=E_R\pm iE_I$, we have 
\begin{eqnarray}
PT|R_{\pm} \rangle=|R_{\mp} \rangle,\quad
\langle L_{\pm}|TP=\langle L_{\mp}|.
\label{H23}
\end{eqnarray}
with time dependencies $|R_{\pm} \rangle\sim \exp(-iE_{\pm}t)=\exp(-iE_Rt\pm E_It)$, $\langle L_{\pm}|=\langle R_{\pm}|V \sim \exp(iE_{\mp}t)=\exp(iE_Rt \pm E_It)$.  Given (\ref{H3}) and (\ref{H4}), we see that these eigenvectors have no overlap with the eigenvectors associated with purely real eigenvalues. In the complex conjugate energy eigenvalue sector we can set  $\sum_n[|R^{+}_{n}\rangle\langle L^{-}_{n}|+|R^{-}_{n}\rangle\langle L^{+}_{n}|]=I$ as summed over however many complex conjugate pairs there are. Also we can set  $\langle L^{-}_{n}|R^{+}_{m}\rangle=\langle L^{+}_{n}|R^{-}_{m}\rangle=\delta (n,m)$, while the previous spectral decomposition of the Hamiltonian given by $H=\sum_n|R_{n}\rangle E_{n}\langle L_{n}|$ is augmented with $H=\sum_n[|R^{+}_{n}\rangle E^{+}_{n}\langle L^{-}_{n}|+|R^{-}_{n}\rangle E^{-}_{n}\langle L^{+}_{n}|]$. Thus just as in our earlier discussion of the decay of some generic state $|A\rangle$ into some generic state $|B\rangle$, the non-trivial overlaps are always between states with exponentially decaying and exponentially growing  behavior in time.

Now while the Hamiltonian does not link the real and complex conjugate sectors the scalar field can. In this mixed sector, with summations being suppressed, the decomposition of the scalar field is given by
\begin{eqnarray}
\phi&=&|R_{i} \rangle \phi_{i-}\langle L_{-}|+|R_{i} \rangle \phi_{i+}\langle L_{+}|
\nonumber\\
&+&|R_{-} \rangle \phi_{-i}\langle L_{i}|+|R_{+} \rangle \phi_{+i}\langle L_{i}|,
\nonumber\\
PT\phi TP&=&|R_{i} \rangle \phi^*_{i-}\langle L_{+}|+|R_{i} \rangle \phi^*_{i+}\langle L_{-}|
\nonumber\\
&+&|R_{+} \rangle \phi^*_{-i}\langle L_{i}|+|R_{-} \rangle \phi^*_{+i}\langle L_{i}|,
\label{H24}
\end{eqnarray}
with $PT\phi TP=\phi$  thus yielding
\begin{eqnarray}
&&\phi_{i-}=\phi^*_{i+},~~ \phi_{i+}=\phi^*_{i-},~~ \phi_{-i}=\phi^*_{+i},~~ \phi_{+i}=\phi^*_{-i},
\nonumber\\
&&\langle L_{i}|\phi |R_{+}\rangle =\phi_{i-},~~  \langle L_{i}|\phi |R_{-}\rangle =\phi_{i+},
\nonumber\\
&&\langle L_{+}|\phi |R_{i}\rangle =\phi_{-i},~~  \langle L_{-}|\phi |R_{i}\rangle =\phi_{+i}.
\label{H25}
\end{eqnarray}
The contribution of this sector to the two-point function is given by 
\begin{eqnarray}
&&\langle \Omega_{\rm L}|\phi(0,t)\phi(0,0)|\Omega_{\rm R}\rangle  e^{-iE_0t}
\nonumber\\
&&=\phi_{0-}\phi_{+0} e^{-iE_{R}t+E_It}+\phi_{0+}\phi_{-0} e^{-iE_{R}t-E_It}.
\label{H26}
\end{eqnarray}
Via (\ref{H25}) we see that the Euclidean time path integral is completely real, just as desired.  
 
On comparing (\ref{H26}) with (\ref{H22}), we see that (\ref{H26})  is a direct continuation of (\ref{H22}),  with pairs of states with real energy eigenvalues in (\ref{H22}) continuing into pairs of states with complex conjugate energy eigenvalues in (\ref{H26}). This pattern is identical to the one exhibited by the two-dimensional matrix example given in (\ref{H1}). Since we have to go through a Jordan-block phase in order to make the continuation from real to complex energy eigenvalues, we can infer that also in the $PT$-symmetric Jordan-Block case the Euclidean time path integral will be real. In fact this very situation has already been observed in a specific model, the quantum-mechanical fourth-order Pais-Uhlenbeck oscillator model. This model is based on the acceleration-dependent Lagrangian  ${{\cal L}}=(\gamma/2)[\ddot{z}^2-(\omega_1^2+\omega^2_2)\dot{z}^2+\omega_1^2\omega_2^2 z^2]$ for the coordinate $z(t)$, where $\omega_1^2$ and $\omega^2_2$ are both real, and $\gamma$ is a positive constant. The Hamiltonian of the theory is $PT$ symmetric, and in the equal-frequency limit becomes Jordan block. For both the unequal-frequency case and the equal-frequency case the Euclidean time path integral is found to be real \cite{Mannheim2007}, with the unequal-frequency path integral continuing into the equal-frequency path integral in the limit,  while nicely generating none other than the non-stationary $\tau e^{-\omega \tau}$ wave function described earlier.

In the event that all the energy eigenvalues of the theory are in complex conjugate pairs, we can calculate two-point functions taken in these states. Since the Hamiltonian does not  induce  transitions  between differing pairs we only need to consider one such pair. In this sector we can expand $\phi$ according to 
\begin{eqnarray}
\phi&=&|R_{+} \rangle \phi_{+-}\langle L_{-}|+|R_{-} \rangle \phi_{-+}\langle L_{+}|,
\nonumber\\
PT\phi TP&=&|R_{-} \rangle \phi^*_{+-}\langle L_{+}|+|R_{+} \rangle \phi^*_{-+}\langle L_{-}|,
\label{H27}
\end{eqnarray}
with $PT\phi TP=\phi$  thus yielding
\begin{eqnarray}
\phi_{+-}=\phi^*_{-+},\qquad \phi_{-+}=\phi^*_{+-},
\nonumber\\
\langle L_{-}|\phi |R_{+}\rangle =\phi_{+-},\qquad  \langle L_{+}|\phi |R_{-}\rangle =\phi_{-+}.
\label{H28}
\end{eqnarray}
In this sector we can thus set
\begin{eqnarray}
&&\langle \Omega_{+}|\phi(0,t)\phi(0,0)|\Omega_{-}\rangle  
=\phi_{-+}\phi_{-+} e^{-iE_{R}t-E_It},
\nonumber\\
&&\langle \Omega_{-}|\phi(0,t)\phi(0,0)|\Omega_{+}\rangle  
=\phi_{+-}\phi_{+-} e^{-iE_{R}t+E_It}.
\label{H29}
\end{eqnarray}
From (\ref{H28}) we see that  the Euclidean time path integral associated with $\langle \Omega_{+}|\phi(0,t)\phi(0,0)|\Omega_{-}\rangle +\langle \Omega_{-}|\phi(0,t)\phi(0,0)|\Omega_{+}\rangle$ is completely real. To summarize, we see that in all the possible cases, we find that if the Hamiltonian is $PT$ symmetric the Euclidean time path integral is real.

To prove the converse, we note  that when we continue the path integral to Euclidean time and take the  large $\tau=it$ limit, the leading term is of the form $\exp(-E_0 \tau)$ where $E_0$ is the energy of the ground state. The next to leading term is the first excited state and so on (as sequenced according to the real parts of the energy eigenvalues). If the Euclidean time path integral is real, it is not possible for there to be any single  isolated complex energy eigenvalue. Rather, any such complex eigenvalues must come in complex conjugate pairs, and likewise for the right/left overlap matrix elements of $\phi(0,0)$. Thus if the Euclidean time path integral is real we can conclude that all the energies and matrix elements are real or appear in complex conjugate pairs. Hence, according to our previous discussion, the Hamiltonian of the theory must be $PT$ symmetric. We thus establish that  $PT$ symmetry is a both necessary and sufficient condition for the reality of the Euclidean time path integral, and  generalize to field theory the analogous result for $|H-\lambda I|$ that was obtained in matrix mechanics.

\section{Constraining the Path Integral Action via $PT$ Symmetry}

The discussion given above regarding path integrals was based on starting with matrix elements of products of quantum fields and rewriting them as path integrals. Thus we begin with the q-number theory in which the quantum-mechanical Hilbert space is already specified and construct a c-number path integral representation of its Green's functions from it. However, if one wants to use path integrals to quantize a theory in the first place one must integrate the exponential of $i$ times the classical action over classical paths. Thus we start with the classical action, and if we have no knowledge beforehand of the structure of the quantum action, we cannot construct the classical action by taking the quantum action and replacing each q-number quantity in it by a c-number (i.e. by replacing q-number operators that obey non-trivial $\hbar$-dependent commutation relations by c-number quantities for which all commutators are zero.) Moreover, while a quantum field theory may be based on Hermitian operators, such Hermiticity is an intrinsically quantum-mechanical concept that cannot even be defined until a quantum-mechanical Hilbert space has been constructed on which the quantum operators can then act. Or stated differently, since path integration is an entirely classical procedure involving integration of a purely classical action over classical paths there is no reference to any Hermiticity of operators in it at all. And even if one writes the Lagrangian in the classical action as the Legendre transform of the classical Hamiltonian, one cannot attach any notion of Hermiticity to the classical  Hamiltonian either.

To try to get round this problem one can argue that since the eigenvalues of Hermitian operators are real, and since such eigenvalues are c-numbers, one should build the classical action out of these eigenvalues, with the classical action then being a real c-number. And if the classical action is real, in Euclidean time $i$ times the action would be real too. The simplest example of a real classical action  is the one inferred from the quantum Lagrangian $m\dot{x}^2/2$ for a free, non-relativistic quantum particle with a q-number position operator that obeys $[x,p]=i\hbar$. On setting $\hbar=0$ one constructs  the classical Lagrangian as the same  $m\dot{x}^2/2$ except that now $x$ is a c-number that obeys $[x,p]=0$. Another familiar example is the scalar field Lagrangian $\partial_{\mu}\phi\partial^{\mu}\phi$, with the same form serving in both the q-number and c-number cases. If we take the fields to be charged, while we could use a Lagrangian of the form  $\partial_{\mu}\phi\partial^{\mu}\phi^*$ in the c-number cease, in the q-number case we would have to use $\partial_{\mu}\phi\partial^{\mu}\phi^{\dagger}$.

Despite this, this prescription fails as soon as one couples to a gauge field. Specifically, one can take the quantum-mechanical $A_{\mu}$ to be Hermitian and the  classical-mechanical $A_{\mu}$ to be real. With such a real $A_{\mu}$ one could introduce a classical Lagrangian of the form $(\partial_{\mu}\phi-A_{\mu}\phi)(\partial^{\mu}\phi^*-A^{\mu}\phi^*)$. Now this particular classical Lagrangian  is not acceptable as a path integration with it would not produce conventional quantum electrodynamics. Rather, to generate  conventional quantum electrodynamics via path integration one must take the classical Lagrangian to be of the form  $(\partial_{\mu}\phi-iA_{\mu}\phi)(\partial^{\mu}\phi^*+iA^{\mu}\phi^*)$. Now in this particular case we already know the answer since the $(\partial_{\mu}\phi-iA_{\mu}\phi)(\partial^{\mu}\phi^{\dagger}+iA^{\mu}\phi^{\dagger})$ form [or equivalently $(i\partial_{\mu}\phi+A_{\mu}\phi)(-i\partial^{\mu}\phi^{\dagger}+A^{\mu}\phi^{\dagger})$] is the form of the quantum-mechanical Lagrangian. However, that does not tell us what classical action to use for other theories for which the quantum-mechanical action is not known ahead of time. 

To address this issue we need to ask why one should include the factor of $i$ in the quantum-mechanical Lagrangian in the first place. The answer is that in quantum mechanics it is not $\partial_{\mu}$ that is Hermitian. Rather, it is $i\partial_{\mu}$. Then since  $\partial_{\mu}$ is anti-Hermitian one must combine it with some anti-Hermitian function of the Hermitian $A_{\mu}$, hence $iA_{\mu}$. We thus have a mismatch between the quantum and classical theories, since while $\partial_{\mu}$ is real it is not Hermitian. We must thus seek some entirely different rule for determining the classical action needed for path integration, one that does not rely on any notion of Hermiticity at all. That needed different rule is $PT$ symmetry.

Under a $PT$ transformation $A_{\mu}$ is $PT$ even. Thus with $\partial_{\mu}$ being $PT$ odd \cite{footnote5}, we see that the combination $\partial_{\mu}-iA_{\mu}$ is also $PT$ odd. Thus, with $(\partial_{\mu}\phi-iA_{\mu}\phi)(\partial^{\mu}\phi^*+iA^{\mu}\phi^*)$ being $PT$ even, $PT$ symmetry is readily implementable at the level of the action. And moreover, $PT$ symmetry can be implemented not just on one classical path such as the stationary one, it can be implemented on every classical path, stationary or non-stationary. When this is done the resulting quantum theory is $PT$ symmetric in exactly the same way as path integration over classical paths all of which are Lorentz invariant yields a quantum theory that is Lorentz invariant too.

To buttress the need for and use of $PT$ symmetry at the level of the action, we recall that we had shown above that the matter field $T^{\mu\nu}$ is $PT$ symmetric. Now if we introduce a general coordinate invariant action $I_{\rm M}=\int d^4 x(-g)^{1/2}{{\cal L}}_{\rm M}$ for the matter field, then its variation with respect to the spacetime metric is given by  $\delta I_{\rm M}=(1/2)\int d^4 x(-g)^{1/2}T^{\mu\nu}\delta g_{\mu\nu}$. Now as a field the spacetime metric is $PT$ even (it transforms the same way as a flavor singlet, color singlet, spin two, quark-antiquark bound state). Thus, with the matter field $T^{\mu\nu}$ (as defined above  in Sec. II  by this very same variational procedure) being $PT$ even, we see that the total matter action is $PT$ even. Thus regardless of any Hermiticity considerations the matter action must be $PT$ symmetric, and thus we  must take the classical action  needed for path integral quantization to be $PT$ symmetric on every classical path.

In addition, we note that while the matter field energy-momentum tensor is defined as the variation of the matter field action with respect to the spacetime metric, it does not follow that the geometric structure of the matter field action depends on the metric alone, since one can use a geometric connection $\Gamma^{\lambda}_{\phantom{\alpha}\mu\nu}$ more general than the standard Levi-Civita connection $\Lambda^{\lambda}_{\phantom{\alpha}\mu\nu}=(1/2)g^{\lambda\alpha}(\partial_{\mu}g_{\nu\alpha} +\partial_{\nu}g_{\mu\alpha}-\partial_{\alpha}g_{\nu\mu})$. One could for instance introduce a torsion-dependent connection of the form $K^{\lambda}_{\phantom{\alpha}\mu\nu}=(1/2)g^{\lambda\alpha}(Q_{\mu\nu\alpha}+Q_{\nu\mu\alpha}-Q_{\alpha\nu\mu})$ where $Q^{\lambda}_{\phantom{\alpha}\mu\nu}=\Gamma^{\lambda}_{\phantom{\alpha}\mu\nu}-\Gamma^{\lambda}_{\phantom{\alpha}\nu\mu}$ is the antisymmetric part of the connection. Or one could use the modified Weyl connection introduced in \cite{Mannheim2014}, viz. $V^{\lambda}_{\phantom{\alpha}\mu\nu}=-(2/3)ig^{\lambda\alpha}\left(g_{\nu\alpha}A_{\mu} +g_{\mu\alpha}A_{\nu}-g_{\nu\mu}A_{\alpha}\right)$ where $A_{\mu}$ is the electromagnetic vector potential. As shown in \cite{Mannheim2014}, both $K^{\lambda}_{\phantom{\alpha}\mu\nu}$ and $V^{\lambda}_{\phantom{\alpha}\mu\nu}$ transform in the same $PT$ way (viz. $PT$ odd) as $\Lambda^{\lambda}_{\phantom{\alpha}\mu\nu}$,  and thus neither of them modifies the $PT$ structure of the theory. 

Our use of the modified  $V^{\lambda}_{\phantom{\alpha}\mu\nu}$ connection is of interest for another reason. When first introduced by Weyl in an attempt to metricate (geometrize) electromagnetism, the connection was taken to be of the form $W^{\lambda}_{\phantom{\alpha}\mu\nu}=-g^{\lambda\alpha}\left(g_{\nu\alpha}A_{\mu} +g_{\mu\alpha}A_{\nu}-g_{\nu\mu}A_{\alpha}\right)$. Apart from an overall normalization factor, this connection differs from the modified one by not possessing the factor of $i$. Since Weyl was working in  classical gravity, everything was taken to be real, with the  $\partial_{\mu}$ derivative in the Levi-Civita connection  being replaced by $\partial_{\mu}-2A_{\mu}$ in order to generate $W^{\lambda}_{\phantom{\alpha}\mu\nu}$. From the perspective of classical physics the Weyl prescription was the natural one to introduce. However, it turns out that this prescription does not work for fermions, since if the Weyl connection is inserted into the curved space Dirac action as is, it is found to drop out identically \cite{Mannheim2014}, with Weyl's attempt to metricate electromagnetism thus failing for fermions. However, when instead the modified  $V^{\lambda}_{\phantom{\alpha}\mu\nu}$ is inserted into the curved space Dirac action, it is found \cite{Mannheim2014} to precisely lead to minimally coupled electromagnetism with action $\int d^4x i\bar{\psi}\gamma^{\mu}(\partial_{\mu}-iA_{\mu})\psi$ (the $2/3$ factor in $V^{\lambda}_{\phantom{\alpha}\mu\nu}$ serves to give $A_{\mu}$ the standard minimally coupled weight). Thus the geometric prescription that leads to the correct coupling of fermions to the vector potential is not to replace $\partial_{\mu}$ by $\partial_{\mu}-2A_{\mu}$ in the Levi-Civita connection, but to replace it by $\partial_{\mu}-(4i/3)A_{\mu}$ instead. We note that it is this latter form that respects $PT$ symmetry, and in so doing it leads to  a geometrically-generated electromagnetic Dirac action that is automatically $PT$ symmetric. Hence we are again led to a $PT$-symmetric action.

The analysis we have given in this paper shows that both the classical and the quantum actions have to be $PT$ symmetric. Now since it has no connection to electric charge, the spacetime  metric is $C$ even where $C$ denotes charge conjugation. Any principle that would force $T^{\mu\nu}$ to be $C$ even would then produce an action  that is fully $CPT$ invariant. It would thus be of interest to find such a principle \cite{footnote6}. Now since $C$ is not preserved in weak interactions, not only $PT$ but also $C$ would then have to broken spontaneously.

\section{Continuing the $PT$ Operator into the Complex Plane}

As we have seen, there are two different ways to obtain a real  Euclidean time path integral in which all energy eigenvalues are real -- the Hamiltonian could be Hermitian, or the theory could be in the real eigenvalue realization of  a $PT$ symmetric Hamiltonian. Thus one needs to ask how does one determine which case is which. Now while there is no known complete answer to this question as yet, a partial answer was given in \cite{Mannheim2013}, one that merits further study. Specifically, the regular (non-Euclidean) real time path integral was studied in some specific models, and it was found that in the Hermitian case the path integral exists with a real measure, while in the $PT$ case the fields in the path integral measure (but not the coordinates on which they depend) needed to be continued into the complex plane. (Continuing the path integral measure into the complex plane is also encountered in 't Hooft's study of quantum gravity \cite{tHooft2011}.) In fact, should this partial answer prove to be a general rule, it would then explain how quantum Hermiticity arises in a purely c-number based path integral quantization procedure in the first place, since the path integral itself makes no reference to any  Hilbert space whatsoever. The general rule would then be that  only if the real time path integral exists with a real measure, and its Euclidean time continuation is real, would the quantum matrix elements that the path integral describes then be associated with a Hermitian Hamiltonian acting on a Hilbert space with a standard Dirac norm.

To see what specifically happens in the non-Hermitian but $PT$-symmetric case, consider the Pais-Uhlenbeck oscillator. Its real time path integral is given by 
\begin{eqnarray}
&&G(z_f,t_f;z_i,t_i)=\int_i^f {{\cal D}}[z,\dot{z}] 
\nonumber\\
&&\times \frac{\gamma}{2}\int_i^fdt\bigg{(}
\ddot{z}^2-(\omega_1^2+\omega^2_2)\dot{z}^2+\omega_1^2\omega_2^2 z^2\bigg{)},
\label{H30}
\end{eqnarray}
with the path integration needing to be over separate $z(t)$ and $\dot{z}(t)$ paths \cite{Mannheim2007} since the equations of motion are fourth-order derivative equations. To enable the path integration  to be asymptotically damped we use the Feynman prescription and replace $\omega_1^2$ and $\omega_2^2$ by $\omega_1^2-i\epsilon$ and $\omega_2^2-i\epsilon$. This then generates an additional contribution to the path integral action of the form 
\begin{eqnarray}
&&i\Delta S =\frac{\gamma}{2}\int_i^f dt \bigg{(}
-2\epsilon\dot{z}^2+\epsilon(\omega_1^2+\omega_2^2) z^2\bigg{)}.
\label{H31}
\end{eqnarray}
While this term provides damping for real $\dot{z}$, it does not do so for real $z$. Rather, in analog to the discussion of the divergent Gaussian wave function given above,  $z$ needs to be continued into the complex plane, with the complex $z$-plane splitting up into regions known as Stokes wedges \cite{Bender2007}. In terms of the shape of the letter $X$,  the $z$ integration converges in its north and south quadrants (a region that includes the imaginary $z$ axis), while the $\dot{z} $ integration converges in analog east and west  quadrants (a region that includes the real $\dot{z}$ axis). The arms of the letter $X$ are known as Stokes lines,  with it being necessary to continue $z$ into the complex plane until it crosses a Stokes line in order to get a well-defined real time path integral. For the Pais-Uhlenbeck oscillator this well-defined path integral is associated with a $PT$-symmetric Hamiltonian and not a Hermitian one, with all energy eigenvalues being real and bounded from below \cite{Bender2008a}, and with the Euclidean time path integral being real and finite \cite{footnote7}. The general rule would thus appear to be that the quantum Hamiltonian associated with both a well-defined real time path integral and a well-defined Euclidean time path integral is Hermitian if the convergent Stokes wedges for the real time path integral include the real axis (in which case the quantization is completely standard), and is $PT$-symmetric instead if they do not. 

In a continuation into the complex plane we need to ask what happens to the $PT$ operator. As we now show, it is continued too so that the $[PT,H]=0$ commutator remains intact.  We give the discussion for particle mechanics, with the generalization to fields being direct.

In classical mechanics one can make symplectic transformations that preserve Poisson brackets. A general discussion may for instance be found in \cite{Mannheim2013}, and we adapt that discussion here and consider the simplest case, namely that of a phase space consisting of just one $q$ and one $p$. In terms of the two-dimensional column vector $\eta=\widetilde{(q,p)}$ and an operator $J=i\sigma_2$ we can write a general Poisson bracket as  
\begin{eqnarray}
\{u,v\}=\frac{\partial u}{\partial q}\frac{\partial v}{\partial p}-\frac{\partial u}{\partial p}\frac{\partial v}{\partial q}
=\widetilde{\frac{\partial u}{\partial \eta}}J\frac{\partial v}{\partial \eta}.
\label{H32}
\end{eqnarray}
If we now make a phase space transformation to a new two-dimensional vector $\eta^{\prime}=\widetilde{(q^{\prime},p^{\prime})}$ according to 
\begin{eqnarray}
M_{ij}=\frac{\partial \eta^{\prime}_i}{\partial \eta_j},\qquad \frac{\partial v}{\partial \eta}=\tilde{M}\frac{\partial v}{\partial \eta^{\prime}},\qquad \widetilde{\frac{\partial u}{\partial \eta}}=\widetilde{\frac{\partial u}{\partial \eta^{\prime}}}M,
\label{H33}
\end{eqnarray}
the Poisson bracket then takes the form
\begin{eqnarray}
\{u,v\}=\widetilde{\frac{\partial u}{\partial \eta^{\prime}}}MJ\tilde{M}\frac{\partial v}{\partial \eta^{\prime}}.
\label{H34}
\end{eqnarray}
The Poisson bracket will thus be left invariant for any  $M$ that obeys the symplectic symmetry relation $MJ\tilde{M}=J$.

In the two-dimensional case the relation $MJ\tilde{M}=J$ has a simple solution, viz. $M=\exp(-i\omega \sigma_3)$, and thus for any $\omega$ the Poisson bracket algebra is left invariant. With $q$ and $p$  transforming as 
\begin{eqnarray}
\eta^{\prime}=e^{-i\omega \sigma_3}\eta,~~ q\rightarrow q^{\prime}=e^{-i\omega}q,~~ p\rightarrow p^{\prime}=e^{i\omega}p,
\label{H35}
\end{eqnarray}
the $qp$ product and the phase space measure $dqdp$ respectively transform into $q^{\prime}p^{\prime}$ and $dq^{\prime}dp^{\prime}$. With the classical action $\int dt (p\dot{q}-H(q,p))$ transforming into $\int dt (p^{\prime}\dot{q}^{\prime}-H(q^{\prime},p^{\prime}))$, under a symplectic transformation the path integral of the theory is left invariant  too.

Now though it is not always stressed in classical mechanics studies, the Poisson bracket algebra is left invariant even if, in our notation, $\omega$ is not pure imaginary, since $i\omega$  is just a pure number. This then permits us to invariantly continue the path integral into the complex $(q,p)$ plane. Now one ordinarily does not do this because one ordinarily works with path integrals that are already well-defined with real $q$ and $p$. However, in the $PT$ case  the path integral is often not well-defined for real $q$ and $p$ but can become so in a suitable Stokes wedge region in the complex $(q,p)$ plane. This means that as one makes the continuation one crosses a Stokes line, with the theories on the two sides of the Stokes line being inequivalent. 

As regards what happens to a $PT$ transformation when we continue into the complex plane, we first need to discuss the effect of $PT$ when $q$ and $p$ are real. When they are real, $P$ effects $q \rightarrow -q$, $p \rightarrow -p$, and $T$ effects $q \rightarrow q$, $p \rightarrow -p$. We can thus set  $PT=-\sigma_3K$ where $K$ effects complex conjugation on anything other than the real $q$ and $p$ that may stand to the right, and set 
\begin{eqnarray}
PT\eta=-\sigma_3\eta.
\label{H36}
\end{eqnarray}
Let us now make a symplectic transformation to a new $PT$ operator $(PT)^{\prime}=MPTM^{-1}$. With $i\omega$ being complex the transformation takes the form
\begin{eqnarray}
MPTM^{-1}=e^{-i\omega\sigma_3}(-\sigma_3)e^{-i\omega^*\sigma_3}K.
\label{H37}
\end{eqnarray}
With $\eta$ being real, we thus obtain  
\begin{eqnarray}
(PT)^{\prime}\eta^{\prime}=e^{-i\omega\sigma_3}(-\sigma_3)e^{-i\omega^*\sigma_3}e^{i\omega^*\sigma_3}\eta=-\sigma_3\eta^{\prime}.~
\label{H38}
\end{eqnarray}
Thus the primed variables transform the same way under the transformed PT operator as the unprimed variables do under the unprimed PT operator. With the Hamiltonian transforming as $H^{\prime}(q^{\prime},p^{\prime})=MH(q,p)M^{-1}$, the $[PT,H]=[(PT)^{\prime},H^{\prime}]=0$ commutator is left invariant, in much the same manner as discussed in \cite{footnote1}. The utility of this remark is that once the path integral is shown to be $PT$ symmetric for all real paths, the $PT$ operator will transform in just the right way to enable the path integral to be $PT$ symmetric for complex paths as well. $PT$ symmetry can thus be used to constrain complex plane path integrals in exactly the same way as it can be used to constrain real ones, and to test for $PT$ symmetry one only needs to do so for the real measure case.

It is also of interest to discuss the quantum analog. Consider a pair of quantum operators $q$ and $p$ that obey $[q,p]=i$. Apply a similarity transformation of the form $\exp(\omega pq)$ where $\omega$ is a complex number. This yields 
\begin{eqnarray}
q^{\prime}&=&e^{\omega pq}q e^{-\omega pq}=e^{-i\omega }q,
\nonumber\\
p^{\prime}&=&e^{\omega pq}p e^{-\omega pq}=e^{i\omega }p,
\label{H39}
\end{eqnarray}
and preserves the commutation relation according to $[q^{\prime},p^{\prime}]=i$. Now introduce quantum operators $P$ and $T$ that obey $P^2=I$, $T^2=I$, $[P,T]=0$, and effect 
\begin{eqnarray}
PqP&=&-q,\qquad TqT=q,\qquad PTqTP=-q,
\nonumber\\
PpP&=&-p,\quad TpT=-p,\qquad PTpTP=p.
\label{H40}
\end{eqnarray}
Under the similarity transformation the $PT$ and $TP$ operators transform according to 
\begin{eqnarray}
(PT)^{\prime}&=&e^{\omega pq}PT e^{-\omega pq}=e^{\omega pq} e^{\omega^* pq}PT,
\nonumber\\
(TP)^{\prime}&=&e^{\omega pq}TP e^{-\omega pq}=TPe^{-\omega^* pq} e^{-\omega pq}.
\label{H41}
\end{eqnarray}
From (\ref{H40}) and (\ref{H41}) we thus obtain
\begin{eqnarray}
&&(PT)^{\prime}q^{\prime}(TP)^{\prime}=e^{\omega pq} e^{\omega^* pq}PTe^{-i\omega }qTPe^{-\omega^* pq} e^{-\omega pq}
\nonumber\\
&&=e^{\omega pq} e^{\omega^* pq}e^{i\omega^* }(-q)e^{-\omega^* pq} e^{-\omega pq}
\nonumber\\
&&=e^{\omega pq} e^{i\omega^* }e^{-i\omega^*} (-q)e^{-\omega pq}=- e^{-i\omega }q=-q^{\prime},
\label{H42}
\end{eqnarray}
\begin{eqnarray}
&&(PT)^{\prime}p^{\prime}(TP)^{\prime}=e^{\omega pq} e^{\omega^* pq}PTe^{i\omega }pTPe^{-\omega^* pq} e^{-\omega pq}
\nonumber\\
&&=e^{\omega pq} e^{\omega^* pq}e^{-i\omega^* }pe^{-\omega^* pq} e^{-\omega pq}
\nonumber\\
&&=e^{\omega pq} e^{-i\omega^* }e^{i\omega^*}p e^{-\omega pq}=e^{i\omega }p=p^{\prime}.
\label{H43}
\end{eqnarray}
Thus the primed variables transform the same way under the transformed PT operator as the unprimed variables do under the unprimed PT operator. With the Hamiltonian being a function of $q$ and $p$, the $[PT,H]=[(PT)^{\prime},H^{\prime}]=0$ commutator is left invariant.

As we see, the classical and quantum cases track into each other as we continue into the complex plane, with both the Poisson bracket and commutator algebras being maintained for every $\omega$. We can thus quantize the theory canonically by replacing Poisson brackets by commutators along any direction in the complex $(q,p)$ plane, and in any such direction there will be a correspondence principle for that direction. With $PT$ symmetry we can thus generalize the notion of correspondence principle to the complex plane. In so doing we see that even if the untransformed $q$ and $p$ are Hermitian, as noted above, the transformed $q^{\prime}$ and $p^{\prime}$ will in general not be since the transformations are not unitary ($(q^{\prime})^{\dagger}=e^{i\omega^*}q^{\dagger}=e^{i\omega^*}q\neq e^{-i\omega}q$). However, what will be preserved is their $PT$ structure, with operators thus having well-defined  transformation properties under a $PT$ transformation.  

In conclusion, we note that $PT$ symmetry is not only more far reaching than Hermiticity, it does not even need to be postulated as it is derivable from a  fundamental principle, namely Poincare invariance. Hermiticity is then just a particular realization of $PT$ symmetry.


\begin{thebibliography}{99}

\bibitem{Bender2007} C.~M.~Bender,~Rep.~Prog.~Phys. {\bf 70}, 947 (2007).

\bibitem{Bender1998} C.~M.~Bender and S.~Boettcher, Phys.~Rev.~Lett.~{\bf 80}, 5243
(1998).

\bibitem{Bender1999} C.~M.~Bender, S.~Boettcher, and P.~N.~Meisinger, J.~Math.~Phys.~{\bf
40}, 2201 (1999).

\bibitem{Special2012} \textit{Special issue on quantum physics with non-Hermitian operators}, C.~Bender,~A.~Fring,~U.~G\"{u}nther,~and~H.~Jones (Guest Editors), J. Phys. A: Math. Theor. \textbf{45}, 444001 - 444036  (2012).

\bibitem{Theme2013} \textit{Theme issue on PT quantum mechanics}, C.~M.~Bender, M.~DeKieviet, and S.~P.~Klevansky (Guest Editors),~Phil.~Trans.~R.~Soc.~A \textbf{371}, issue 1989  (2013).

\bibitem{Mannheim2013} P.~D.~Mannheim,~Phil.~Trans.~R.~Soc.~A \textbf{371}, 20120060 (2013).

\bibitem{Bender2008a} C.~M.~Bender and P.~D.~Mannheim,~Phys.~Rev.~Lett. \textbf{100}, 110422 (2008).

\bibitem{Bender2008b} C.~M.~Bender and P.~D.~Mannheim,~Phys.~Rev.~D \textbf{78}, 025022 (2008).

\bibitem{Mannheim2012} P.~D.~Mannheim,~Found.~Phys. \textbf{42}, 388 (2012).


\bibitem{Bender2002} C.~M.~Bender, M.~V.~Berry, and A.~Mandilara,~J.~Phys.~A: Math.~Gen.~{\bf 35}, L467 (2002).

\bibitem{Bender2010} C.~M.~Bender and P.~D.~Mannheim,~Phys.~Lett.~A \textbf{374}, 1616 (2010).

\bibitem{footnote1} If we set $P=\pi$, $T=\tau K =K \tau^*$  and require that $P^2=I$, $T^2=I$, $[P,T]=0$, we obtain $\pi^2=I$, $\tau\tau^*=I$, $\pi\tau=\tau\pi^*$. If we now make a similarity transform $SPS^{-1}=P^{\prime}$, $STS^{-1}=T^{\prime}$ and set $P^{\prime}=\pi^{\prime}$, $T^{\prime}=\tau^{\prime}K$, then  with $\pi^{\prime}=S\pi S^{-1}$, $\tau^{\prime}=S\tau (S^{-1})^*$ we obtain $P^{\prime 2}=I$, $T^{\prime 2}=I$, $[P^{\prime},T^{\prime}]=0$, and $\pi^{\prime 2}=I$, $\tau^{\prime} \tau^{{\prime}*}=I$, $\pi^{\prime}\tau^{\prime}=\tau^{\prime}\pi^{{\prime}*}$. If we transform a Hamiltonian $H$ obeying $H=PTHTP=\pi\tau H^*\tau^*\pi$, we find that $H^{\prime}=SHS^{-1}=P^{\prime}T^{\prime}H^{\prime}T^{\prime}P^{\prime}=\pi^{\prime}\tau^{\prime} H^{{\prime} *}\tau^{{\prime}*}\pi^{\prime}$, with $PT$ symmetry being maintained.

\bibitem{footnote2} An analogous situation is found in differential geometry. To test whether or not an arbitrary spacetime metric $g_{\mu\nu}$ can be brought to a flat form $\eta_{\mu\nu}$ by a coordinate transformation, one evaluates the Riemann tensor that is derived from it. If the Riemann tensor is non-zero then the metric is not coordinate equivalent to flat.

\bibitem{footnote3} In \cite{Bender2010} it was also shown that for diagonalizable matrices one can always construct an operator $C$ that obeys $[C,H]=0$, $C^2=I$.  If in addition $C$ obeys $[C,PT]=0$ then all the eigenvalues of $H$ are real, and if $[C,PT]\neq 0$ then the eigenvalue spectrum includes complex conjugate pairs. If no $C$ exists, the Hamiltonian is Jordan block. Thus for matrices the conditions $[H,PT]=0$, $[C,PT]=0$ are necessary and sufficient for Hermiticity.

\bibitem{Mannheim2014} P.~D.~Mannheim,~\textit{$PT$  Symmetry, Conformal Symmetry, and the Metrication of Electromagnetism}, arXiv:1407.1820 [hep-th], July 2014.

\bibitem{Callan1970} C.~G.~Callan,~S. Coleman,~and~R.~Jackiw,~Anns.~Phys.~\textbf{59}, 42 (1970).

\bibitem{footnote4} With $t$-plane singularities having $t_I>0$ (the typical oscillator path integral behaves as $1/\sin[(\omega-i\epsilon)t]$), and with circle at infinity terms vanishing in the lower half plane (cf. $\exp[-i\omega(t_R+it_I)]$), with $\tau=it$ a lower right quadrant Wick rotation yields $i\int_{0}^{\infty} dt =-i\int_{-i\infty}^{0} dt =\int_{0}^{\infty} d\tau$.


\bibitem{Mannheim2007} P.~D.~Mannheim,~Found.~Phys. \textbf{37}, 415 (2007).

\bibitem{footnote5} In a $PT$ transformation on the coordinates, $\partial_{\mu}$ transforms into $-\partial_{\mu}$. In a $PT$ transformation on the fields $\partial_{\mu}\phi(x^{\lambda})$ transforms into $ \partial_{\mu}\phi(-x^{\lambda})$, i.e. into $-[\partial/\partial (-x^{\mu})]\phi(-x^{\lambda})$.  Thus, under a $d^4x $ integration the $PT$ transform of $\partial_{\mu}\phi(x^{\lambda})$ acts as $-\partial_{\mu}\phi(x^{\lambda})$. Thus, under a transformation on coordinates or fields, in the action  $\partial_{\mu}$ acts as a $PT$ odd operator.



\bibitem{footnote6} A possible principle that would produce $C$ invariance is grandunification, since with it one can treat particles and antiparticles equivalently. In a grandunification based on $SO(10)$ for instance, all the 16 quark, lepton, antiquark, and antilepton members of a common fundamental fermionic family are placed in its 16-dimensional irreducible spinor representation, with their couplings to a 45-dimensional irreducible representation of gauge bosons being $C$ invariant.  As regards  the $CPT$ theorem, as first introduced,  it enabled one to constrain the possible actions that could be used in quantum field theory without needing to have to specify any particular one. If grandunification is correct it then does specify a very specific action for the strong, electromagnetic and weak forces, one that then is $CPT$ invariant. Moreover, with $g_{\mu\nu}$ itself being $CPT$ even, $CPT$ symmetry is maintained if one couples a grandunified theory to gravity. 

\bibitem{tHooft2011} G.~'t Hooft,~Found.~Phys.~\textbf{41}, 1829 (2011).

\bibitem{footnote7} In Euclidean time the Pais Uhlenbeck Lagrangian takes the form ${{\cal L}}=(\gamma/2)[(d^2z/d\tau^2)^2+(\omega_1^2+\omega_2^2)(dz/d\tau)^2+\omega_1^2\omega_2^2z^2]$. On putting $z$ on the imaginary axis as per (\ref{H31}), with positive $\gamma$ the Lagrangian is  negative definite in every Euclidean path. With the needed Euclidean action being given by $\int d\tau{{\cal L}}$ as per \cite{footnote4},  the needed action is negative definite on every Euclidean path, and the Euclidean time path integral is finite.

\end{thebibliography}
\end{document}